\documentclass[cameraready]{Interspeech}

\usepackage{graphicx}
\usepackage{amsmath,graphicx,hyperref,amssymb,multirow,makecell,float}
\title{Blind Room Impulse Response Identification via Reverberant Speech Spectrum Reconstruction}

\author[affiliation={1,2}, orcid=0000-0001-5768-0658]{Pengyu}{Wang}
\author[affiliation={2,3}, correspondingauthor, orcid=0000-0003-0393-9905]{Xiaofei}{Li}

\address{
    $^1$ Zhejiang University, Hangzhou, China \\
    $^2$ Westlake University, Hangzhou, China \\
    $^3$ Westlake Institute for Advanced Study, Hangzhou, China
}

\email{wangpengyu@westlake.edu.cn, lixiaofei@westlake.edu.cn}


\begin{document}

\maketitle

\begin{abstract}
This paper proposes Rec-RIR for blind room impulse response (RIR) identification.
Based on the convolutive transfer function (CTF) approximation, we propose a multi-task deep neural network, which sequentially removes noise and reverberation from speech recording, and estimates the CTF filter by reverberant speech spectrum reconstruction.
Subsequently, a pseudo intrusive measurement process is employed to convert the CTF filter into RIR by simulating a common intrusive RIR measurement procedure.
Experimental results demonstrate that Rec-RIR achieves state-of-the-art (SOTA) performance in blind RIR identification. 
\end{abstract}

\section{Introduction}

Room impulse response (RIR) models the complete propagation process of sound and contains key information of the acoustic environment.
RIR estimation facilitates a wide range of applications, such as speech enhancement \cite{vincent2018audio}, speech recognition \cite{ratnarajah2023towards} and augmented/virtual reality \cite{schissler2014high}.
A common intrusive method for measuring RIR is to play a known excitation signal, such as a maximum length sequence (MLS) or a sine sweep, and perform inverse filtering on the observation \cite{stan2002comparison,jalmby2023low}.
However, intrusive measurement is expensive and not always practical in real-world applications.
In contrast, blind RIR identification aims to estimate RIRs from observations of a specific form (e.g., single-channel speech recordings in this work) without relying on known source signals (e.g., source speech). 

Deep neural network (DNN)-based blind RIR identification uses data-driven methods to directly learn the mapping from observed signals to RIR-related representations.
Many approaches encode fixed-length observations in the time domain and use a decoder to obtain fixed-length RIR estimates.
For example, S2IR-GAN \cite{ratnarajah2023towards} uses a generative adversarial network (GAN) to estimate RIRs up to 0.25~s.
However, estimating RIRs in the time domain remains challenging due to their long taps, which typically consist of thousands of samples.
In FiNS \cite{steinmetz2021filtered}, the RIR is decomposed into three distinct components: the direct-path impulse, early reflections, and decaying filtered noise signals. 
A time-domain encoder-decoder architecture is employed to learn both the early RIR and the noise filter, which significantly reduces the output dimension, thus alleviating the difficulty of estimating long RIRs.
SG-RIR \cite{liao23_interspeech} employs a segmental network that generates one segment of the complete RIR each time, with the network architecture shared across all segments.
According to the convolutive transfer function (CTF) approximation~\cite{talmon2009relative}, the reverberation effect can be modeled by narrow-band filtering in the short-time Fourier transform (STFT) domain.
Therefore, the estimation of a long RIR can be realized by estimating a much shorter CTF filter, thus simplifying the task.
For instance, BUDDy~\cite{11024065} and VINP~\cite{wang2025vinp} utilize maximum likelihood estimation to iteratively estimate CTF filters, and achieve high-precision blind RIR identification.

This work proposes a blind \textbf{RIR} identification network based on reverberant speech spectrum \textbf{Rec}onstruction, named \textbf{Rec-RIR}.
Based on the CTF approximation, we design a novel multi-task DNN that extracts and fuses embeddings of both reverberant and clean speech to directly estimate CTF filters.
Subsequently, the pseudo intrusive measurement process proposed in VINP \cite{wang2025vinp} is adopted to convert the estimated CTFs into RIRs.
Distinct from most existing methods, Rec-RIR enables long RIR modeling with a maximum duration of 0.96 s, eliminates the need for iterative computation, and achieves superior performance over state-of-the-art (SOTA) baselines.
The main contributions of Rec-RIR are summarized as follows:
a) To the best of our knowledge, this is the first work that formulates blind RIR identification as a supervised reverberant speech spectrum reconstruction task,
which enables effective and accurate estimation of long CTF filters and RIRs;
b) We propose a novel multi-task DNN that sequentially removes noise and reverberation from observations, and fuses the features of reverberant and clean speech to accomplish CTF estimation;
c) Rec-RIR delivers consistent SOTA performance in both acoustic parameter estimation and RIR waveform estimation.
Codes are available at https://github.com/Audio-WestlakeU/Rec-RIR.

\section{Signal Model}
\label{sec:signal_model}
Considering a single speaker and a single microphone in the space, the observation can be modeled in the time domain as
\begin{equation}
    y(n)=x(n)+w(n)=h(n)*s(n)+w(n),
\end{equation}
where $n$ is the sample index, $x(n)$ and $w(n)$ denote reverberant speech and additive noise, respectively.
$x(n)$ is the convolution of clean speech $s(n)$ and RIR $h(n)$. 
Note that we constrain $s(n)$ to have the delay and gain of the direct-path observation, and constrain $h(n)$ to start with the direct-path impulse, to obtain a unique solution for RIR identification.
In the STFT domain, according to the CTF approximation \cite{talmon2009relative}, there is
\begin{equation}
\begin{aligned}
        Y(f,t)&=X(f,t)+W(f,t)\\
        &\approx \sum\limits_{l=0}^{L-1}H_l(f)S(f,t-l)+W(f,t),
\end{aligned}
\end{equation}
where $f\in [1,F]$ and $t\in [1,T]$ are the indices of frequency band and frame, respectively. $Y(f,t)$, $X(f,t)$, $S(f,t)$ and $W(f,t)$ are the STFT coefficients of $y(n)$, $x(n)$, $s(n)$ and $w(n)$, respectively.
$H_l(f)$ is the $l$th coefficient of the $L$-order CTF filter at frequency band $f$.
For simplicity, we define $\mathbf{Y},\mathbf{X},\mathbf{S},\mathbf{W}\in\mathbb{C}^{F\times T}$ as the spectra of observation, reverberant speech, clean speech and noise, respectively, and define $\mathbf{H}\in \mathbb{C}^{F\times L}$ as the entire CTF filter.
Based on the above definitions, we have
\begin{equation}
    \label{specsignal}
    \mathbf{Y}=\mathbf{X}+\mathbf{W}\approx\mathbf{H}\circledast\mathbf{S}+\mathbf{W},
\end{equation}
where $\circledast$ denotes convolution along the frame axis.
CTF approximation models the reverberation effect within narrow-band filtering in the STFT domain.
Since the approximation error is sufficiently small to be negligible \cite{avargel2007system,tang2025personal},
RIR and CTF contain nearly identical information.
Thus, we can convert the estimation of the RIR into that of the CTF, thereby shortening the filter and simplifying the problem.

\section{Proposed Method}
\label{sec:method}
\begin{figure}[htbp]
    \centering
    \includegraphics[width=0.85\linewidth]{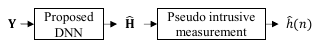}
    \vspace{-0.5em}
    \caption{Workflow of Rec-RIR.}
    \vspace{-0.5em}
    \label{fig:work_flow}
\end{figure}

In Rec-RIR, we propose a multi-task DNN to learn the mapping from the observation $\mathbf{Y}$ to the CTF filter estimate $\hat{\mathbf{H}}$.
Subsequently, a pseudo intrusive measurement process is introduced to convert the CTF filter estimate $\hat{\mathbf{H}}$ into RIR estimate $\hat h(n)$.
The workflow of Rec-RIR is illustrated in Fig.~\ref{fig:work_flow}.

\subsection{Proposed network}
\label{sec:proposed_network}

\begin{figure*}[htbp]
    \centering
    \includegraphics[width=0.95\linewidth]{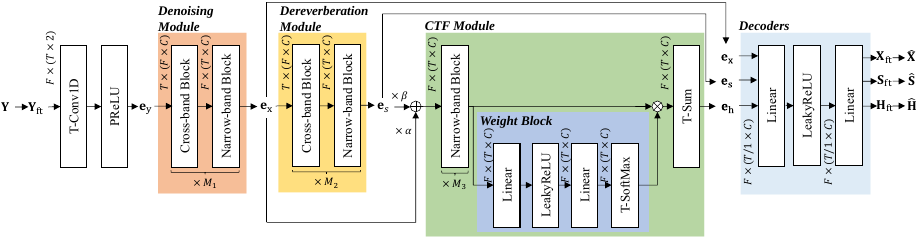}
    \vspace{-1em}
    \caption{Architecture of the proposed DNN.}
    \vspace{-1em}
    \label{fig:model}
\end{figure*}

According to the signal model in Eq.~(\ref{specsignal}), noise acts as an interference term in RIR identification and thus requires elimination. 
Moreover, speech dereverberation and blind RIR identification share an inherent unity. 
That means, accomplishing either task provides explicit guiding information to facilitate the solution of the other. 
Given the above two points, we propose a novel network that extracts and fuses embeddings of both reverberant and clean speech to estimate the CTF filter, whose architecture is illustrated in Fig.~\ref{fig:model}.
The network takes as input the real and imaginary parts of observation $\mathbf Y$, denoted as 
\begin{equation}
    {\mathbf{Y}}_\mathrm{ft}[f,t,:]=[\mathrm{Re}\{Y(f,t)\},\mathrm{Im}\{Y(f,t)\}]\in \mathbb{R}^2.
\end{equation}
After that, a temporal convolution with kernel size of 5 (T-Conv1D) and a PReLU activation \cite{he2015delving} are employed to generate the embedding of observation as
\begin{equation}
    \mathbf{e}_\mathrm{y}[f,t,:]=\mathrm{PReLU}(\mathrm{T\text{-}Conv1D}({\mathbf{Y}}_\mathrm{ft}[f,t,:]))\in \mathbb{R}^C,
\end{equation}
where $C$ is the dimension for embedding $\mathbf{e}_\mathrm{y}$ in each time-frequency (T-F) bin.
Subsequently, we design a denoising module, a dereverberation module, and a CTF module to sequentially remove noise and reverberation in the observation, and finalize CTF estimation.
Their specific structures are as follows.

\subsubsection{Denoising module}
Given embedding $\mathbf{e}_\mathrm{y}$, the denoising module outputs the embedding of the noise-free reverberant speech, denoted as $\mathbf{e}_\mathrm{x}$.
The denoising module is composed of $M_1$ interleaved 'cross-band' and 'narrow-band' blocks, which are mature network blocks and have been proven to be a highly effective structure in single-channel speech denoising \cite{11097896}.

\begin{figure}[htbp]
    \centering
    \includegraphics[width=\linewidth]{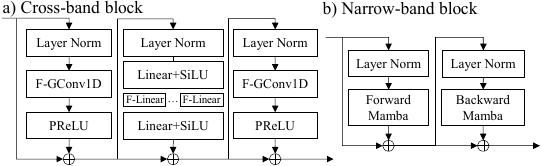}
    \vspace{-2em}
    \caption{Cross-band and narrow-band blocks \cite{wang2025vinp,10423815}.}
    \vspace{-1.2em}
    \label{fig:cnarch}
\end{figure}

The cross-band block processes frames independently and is designed for learning full-band signal dependencies.
We employ the original cross-band block in SpatialNet \cite{10423815}, which comprises cascaded frequency convolutional layers (F-GConv1D), across-frequency linear layers (F-Linear), and a second frequency convolutional layer, with layer normalization and activations \cite{he2015delving,hendrycks2016gaussian}.
The narrow-band block processes frequency bands independently and is designed for learning narrow-band signal dependencies.
We employ the narrow-band block in VINP-oSpatialNet \cite{wang2025vinp}, which comprises cascaded forward and backward Mamba layers with layer normalization.
The forward Mamba layer is a normal Mamba layer \cite{gu2024mamba}.
The backward Mamba layer is identical to the forward one, with the only difference being that its input and output are temporally reversed.
Due to space limitations, the specific structures of the cross-band and narrow-band blocks are not discussed in detail herein, but they can be found in Fig. \ref{fig:cnarch} and our references. 

Denote the function of interleaved $M_1$ cross-band and narrow-band blocks as $f^{M_1}_{\mathrm{cb+nb}}(\cdot)$, the output of denoising module is
\begin{equation}
\mathbf{e}_\mathrm{x}=f^{M_1}_{\mathrm{cb+nb}}(\mathbf{e}_\mathrm{y})\in \mathbb{R}^{F\times T \times C}.
\end{equation}

Moreover, to guide training, a decoder composed of linear layers and LeakyReLU activation is proposed to decode the embedding $\mathbf{e}_\mathrm{x}$ into the estimate of the reverberant speech $\hat{\mathbf{X}}$ as
\begin{equation}
\begin{aligned}
        {\mathbf{X}}_\mathrm{ft}=\mathrm{Linear}(\mathrm{LeakyReLU}(\mathrm{Linear}(\mathbf{e}_\mathrm{x})))\in \mathbb{R}^{F\times T \times 2}
\end{aligned}
\end{equation}
\begin{equation}
    \hat{\mathbf{X}}[f,t]={\mathbf{X}}_\mathrm{ft}[f,t,1]+\mathrm{j}*{\mathbf{X}}_\mathrm{ft}[f,t,2]\in \mathbb{C}^1.
\end{equation}
An auxiliary loss is introduced to guide denoising as
\begin{equation}
\begin{aligned}
\label{eq:loss_x}
    \mathcal{L}_\mathrm{denoi}&=\mathcal{L}_\mathrm{Mag+RI}(\hat{\mathbf{X}},\mathbf{X}),
\end{aligned}
\end{equation}
where
\begin{equation}
\begin{aligned}
    &\mathcal{L}_\mathrm{Mag+RI}(\hat{\mathbf{X}},\mathbf{X})
    =\frac{1}{FT}\left[|||\hat{\mathbf{X}}|-|\mathbf{X}|||_1\right.\\
    &+||\mathrm{Re}\{\hat{\mathbf{X}}\}-\mathrm{Re}\{\mathbf{X}\}||_1
    +\left.||\mathrm{Im}\{\hat{\mathbf{X}}\}-\mathrm{Im}\{\mathbf{X}\}||_1\right].
\end{aligned}
\end{equation}

\subsubsection{Dereverberation module}
Given embedding $\mathbf{e}_\mathrm{x}$, the dereverberation module further removes the reverberation and outputs embedding of clean speech, denoted as $\mathbf{e}_\mathrm{s}$.
Since interleaved cross-band and narrow-band blocks have also been proven to be a highly effective structure in single-channel speech dereverberation \cite{wang2025vinp,11097896},
the dereverberation module shares the same structure as the denoising module, except that there are $M_2$ interleaved blocks.
The output of dereverberation module $f^{M_2}_{\mathrm{cb+nb}}(\cdot)$ is
\begin{equation}
\mathbf{e}_\mathrm{s}=f^{M_2}_{\mathrm{cb+nb}}(\mathbf{e}_\mathrm{x})\in \mathbb{R}^{F\times T \times C}.
\end{equation}

Like the denoising module, the dereverberation module also leverages a decoder to guide its learning, which aims to decode $\mathbf{e}_\mathrm{s}$ into the clean spectrum estimate $\hat{\mathbf{S}}$. 
The decoder shares the same structure as the denoising module, written as 
\begin{equation}
    \mathbf{S}_\mathrm{ft}=\mathrm{Linear}(\mathrm{LeakyReLU}(\mathrm{Linear}(\mathbf{e}_\mathrm{s})))\in \mathbb{R}^{F\times T \times 2},
\end{equation}
\begin{equation}
    \hat{\mathbf{S}}[f,t]=\mathbf{S}_\mathrm{ft}[f,t,1]+\mathrm{j}*\mathbf{S}_\mathrm{ft}[f,t,2]\in \mathbb{C}^1.
\end{equation}
An auxiliary loss is also introduced to guide dereverberation as
\begin{equation}
\begin{aligned}
\label{eq:loss_s}
    \mathcal{L}_\mathrm{dereverb}&=\mathcal{L}_\mathrm{Mag+RI}(\hat{\mathbf{S}},\mathbf{S}).
\end{aligned}
\end{equation}

\subsubsection{CTF module}
This module performs implicit neural inverse filtering for CTF estimation.
Specifically, it takes weighted reverberant speech embedding $\mathbf{e}_\mathrm{x}$ and clean speech embedding $\mathbf{e}_\mathrm{s}$ as input, and outputs CTF embedding $\mathbf{e}_\mathrm{h}$.
According to the CTF approximation, the narrow-band filter banks across frequency bands are independent.
Therefore, $M_{\mathrm{3}}$ stacked narrow-band blocks $f^{M_3}_{\mathrm{nb}}(\cdot)$ are employed to further extract the hidden feature as
\begin{equation}
\mathbf{h}=f^{M_3}_{\mathrm{nb}}(\alpha\mathbf{e}_\mathrm{x}+\beta\mathbf{e}_\mathrm{s})\in \mathbb{R}^{F\times T \times C},
\end{equation}
where $\alpha$ and $\beta$ are learnable.
To process observation of arbitrary length, a weight block is designed to calculate the importance of hidden feature $\mathbf{h}$ at all frames.
The weight block consists of 2 linear layers, a LeakyReLU activation and a SoftMax activation along the frame axis (T-SoftMax), written as
\begin{equation}
\begin{aligned}
&\mathbf{w}[f,:,:]\\
&=\mathrm{T}\text{-}\mathrm{SoftMax}\left(\mathrm{Linear}\left(\mathrm{LeakyReLU}\left(\mathrm{Linear}\left(\mathbf{h}[f,:,:]\right)\right)\right)\right)\\
&\in \mathbb{R}^{T \times C}.
\end{aligned}
\end{equation}
The hidden feature $\mathbf{h}$ is weighted and summarized along the frame axis to generate CTF embedding $\mathbf{e}_\mathrm{h}$ as 

\begin{equation}
    \mathbf{e}_\mathrm{h}[f,:,:]=\sum_{t=1}^{T}\left(\mathbf{h}[f,t,:]\circ\mathbf{w}[f,t,:]\right)\in \mathbb{R}^{1 \times C},
\end{equation}
where $\circ$ denotes Hadamard product.

Finally, a decoder sharing the same structure with the previous two decoders maps $\mathbf{e}_\mathrm{h}$ into CTF estimate $\hat{\mathbf{H}}$ as
\begin{equation}
    \mathbf{H}_\mathrm{ft}=\mathrm{Linear}(\mathrm{LeakyReLU}(\mathrm{Linear}(\mathbf{e}_\mathrm{h})))\in \mathbb{R}^{F\times 1 \times 2L}
\end{equation}
\begin{equation}
    \hat{\mathbf{H}}[f,l]=\mathbf{H}_\mathrm{ft}[f,1,l]+\mathrm{j}*\mathbf{H}_\mathrm{ft}[f,1,L+l]\in \mathbb{C}^1.
\end{equation}
A primary loss is introduced to guide CTF estimation through explicit reconstruction of the reverberant speech spectrum as
\begin{equation}
\label{eq:lctf}
    \mathcal{L}_\mathrm{ctf}=\mathcal{L}_\mathrm{Mag+RI}(\hat{\mathbf{H}}\circledast\mathbf{S},\mathbf{X}).
\end{equation}

\subsection{Multi-task loss function}
As described in Section \ref{sec:proposed_network}, the proposed network has multiple outputs: the estimation of CTF filter serves as the primary output, and the remaining outputs are auxiliary.
For balancing the three losses, we apply weighting to them as
\begin{equation}
\label{eq:losses}
    \mathcal{L}=\mathcal{L}_{\mathrm{ctf}}
    +\lambda_{\mathrm{denoi}}\mathcal{L}_{\mathrm{denoi}}
    +\lambda_{\mathrm{dereverb}}\mathcal{L}_{\mathrm{dereverb}}.
\end{equation}
In this work, we empirically set $\lambda_{\mathrm{denoi}}=\lambda_{\mathrm{dereverb}}=1$.
The role of auxiliary losses is discussed in Section \ref{sec:ablation_study}.
Notice that although clean speech is used for supervised learning, Rec-RIR operates in a fully blind manner at inference time.

In theory, blind RIR identification can be ill-posed due to missing frequency components and/or unconstrained delay and gain of the source signal.
However, speech signals typically provide adequate spectral coverage.
Moreover, in this work, $\mathbf{S}$ and $\mathbf{X}$ in Eq.~(\ref{eq:lctf}) are time- and gain-aligned, which introduces additional constraints.
Therefore, the identification task is sufficiently constrained to enable a data-driven solution.

\subsection{Pseudo intrusive measurement process}
We employ the pseudo intrusive measurement process proposed in \cite{wang2025vinp} to convert the CTF filter to RIR by simulating a common intrusive measurement process.
Given a logarithmic sine sweep as an excitation signal $e(n)$, there is an inverse filter $v(n)$ that satisfies $e(n)*v(n)=\delta(n)$, where $\delta(n)$ is an impulse~\cite{stan2002comparison,farina2000simultaneous}.
Playing such an excitation signal, according to Eq.~(\ref{specsignal}), the measured spectrum is 
$\mathbf{Z}\approx\hat{\mathbf{H}}\circledast\mathbf{E}$,
where $\mathbf{E}$ is the STFT of $e(n)$. 
Finally, the RIR can be estimated by inverse filtering as
$\hat{h}(n)=z(n)*v(n)$,
where $z(n)$ is the inverse STFT of $\mathbf{Z}$.

\section{Experiments}
\label{sec:experiments}
\begin{table*}[ht!]
\centering
\caption{Blind RIR identification results on SimACE}
\label{tab:RIR_estimation}
\vspace{-1em}
\resizebox{0.9\linewidth}{!}{
\begin{tabular}{c|cc|ccc|ccc|ccc}
    \Xhline{1pt}
    \multirow{2}{*}{Method}&\multicolumn{2}{c|}{RIR-50 ms}&\multicolumn{3}{c|}{RT60 (s)}& \multicolumn{3}{c|}{DRR (dB)}& \multicolumn{3}{c}{C50 (dB)}\\
    \Xcline{2-12}{0.4pt}
    &RMSE$\downarrow$&$\bar\rho$$\uparrow$&MAE$\downarrow$&RMSE$\downarrow$&$\rho$$\uparrow$&MAE$\downarrow$&RMSE$\downarrow$&$\rho$$\uparrow$&MAE$\downarrow$&RMSE$\downarrow$&$\rho$$\uparrow$\\
    \Xhline{0.4pt}
    
    FiNS~\cite{steinmetz2021filtered} (2021)&0.067&0.409&0.113&0.167&0.857&2.153&2.639&0.737&6.489&6.597&0.925\\
    BUDDy~\cite{11024065} (2025)&0.057&0.621&0.122&0.166&0.952&3.673&4.360&0.774&4.109&4.477&0.779\\
    VINP-TCN+SA+S \cite{wang2025vinp} (2025)&0.050&0.695&0.089&0.124&0.934&3.256&3.764&0.872&0.914&1.135&0.961\\
    VINP-oSpatialNet \cite{wang2025vinp} (2025)&0.050&0.703&0.103&0.154&0.895&2.398&2.875&0.908&0.977&1.295&0.943\\
    \Xhline{0.4pt}
    \textbf{Rec-RIR (prop.)}&\textbf{0.040}&\textbf{0.805}&\textbf{0.069}&\textbf{0.104}&\textbf{0.994}&\textbf{0.684}&\textbf{0.794}&\textbf{0.994}&\textbf{0.858}&\textbf{1.019}&\textbf{0.978}\\
    \Xhline{1pt}
\end{tabular}
}
\vspace{-0.5em}
\end{table*}

\subsection{Datasets}
We use the 16 kHz training and test sets employed in VINP \cite{wang2025vinp}, which are generated by convolving clean speech with RIRs, followed by noise addition.

In the training set, we use 200~h high-quality clean speech from DNS Challenge~\cite{reddy2020interspeech}, VCTK~\cite{valentini2016investigating}, and EARS~\cite{richter2024ears}.
We use 100,000 reverberant and direct-path RIR pairs simulated by gpuRIR~\cite{diaz2021gpurir}, in which the speaker and microphone are randomly placed in rooms with dimensions randomly selected within a range of 3 m to 15 m for length and width, and 2.5 m to 6 m for height.
Reverberant RIRs have RT60s uniformly distributed within the range of 0.2~s to 1.5~s. 
Correspondingly, direct-path RIRs are generated using
an absorption coefficient of 0.99.
Noise comes from NOISEX-92~\cite{varga1993assessment} and the training set of REVERB Challenge~\cite{kinoshita2013reverb}.
The signal-to-noise ratio (SNR) is uniformly distributed within the range of 5~dB to 20~dB.

We use the SimACE test set proposed in \cite{wang2025vinp}, in which the recordings are generated by convolving the clean speech from WSJ0 corpus~\cite{paul1992design} with the downsampled and measured RIRs from 'single' subset of ACE Challenge~\cite{eaton2016estimation}, and adding noise from the test set in REVERB Challenge with a SNR of 20 dB.

\subsection{Settings}
Before Rec-RIR, the input recordings are normalized by their maximum absolute value in time domain.
The STFT analysis and synthesis windows are square-root Hann windows with a length of 512 samples and 50\% overlap, which means $F=257$.
The hyperparameters of Rec-RIR are set to $M_{\mathrm{1}}=2$, $M_{\mathrm{2}}=6$, $M_{\mathrm{3}}=6$ and $C=96$.
The CTF length is set to $L=60$, corresponding to RIR with an effective length of 0.96 s.
With the above settings, Rec-RIR has 3.1 M parameters and a computational complexity of 35.2 GMACs/s.

For training, we segment speech utterances into 4 s and use 97,092 samples per epoch with a batch size of 4. 
We adopt the AdamW optimizer~\cite{loshchilov2017decoupled}.
The learning rate restarts at 0.001 and decays following a cosine schedule in each epoch.

In the pseudo intrusive measurement process, we employ the logarithmic sine sweep signal~\cite{stan2002comparison,farina2000simultaneous} as that in VINP \cite{wang2025vinp}.

\subsection{Comparison methods and metrics}
Comparison methods include advanced approaches FiNS~\cite{steinmetz2021filtered}, BUDDy~\cite{11024065} and VINP~\cite{wang2025vinp}.
Part of convolutional kernels in FiNS are modified to handle the 16~kHz inputs.
We evaluate them using the estimation accuracy of acoustic parameters, including reverberation time (RT60), direct-to-reverberant ratio (DRR), and clarity (C50).
We present their mean absolute error (MAE), root mean square error (RMSE), and Pearson correlation coefficient ($\rho$).
Also, we provide RMSE and average $\rho$ ($\bar\rho$) of early reflections, corresponding to the 50~ms duration after the direct-path impulse in RIR, marked as RIR-50~ms.

\subsection{Results and analysis}
As shown in Table~\ref{tab:RIR_estimation}, Rec-RIR achieves the highest accuracy and consistency (evidenced by high Pearson coefficients) in estimating both acoustic parameters and early reflections, significantly outperforming baselines by a clear margin.
Among the comparison methods, FiNS is designed for fixed-length inputs and is therefore limited in exploiting long observations, while BUDDy and VINP rely on iterative optimization.
In contrast, the task formulation and architecture of Rec-RIR allow end-to-end estimation of CTF filters from inputs of arbitrary length, leading to better identification performance.

\begin{figure}[ht!]
    \centering
    \includegraphics[width=\linewidth]{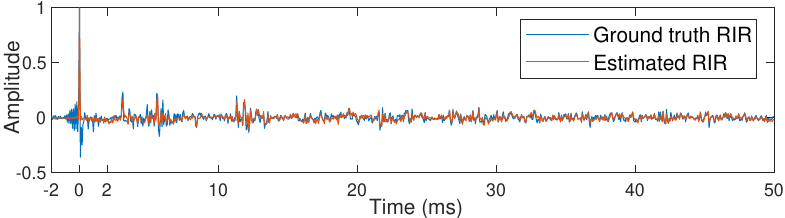}
    \vspace{-2em}
    \caption{Example of estimated RIR (normalized).}
    \vspace{-0.5em}
    \label{fig:RIR}
\end{figure}

Fig.~\ref{fig:RIR} shows an estimated RIR whose RT60 is 1.26~s.
As observed, Rec-RIR is able to accurately estimate the early reverberation, including the first reflection impulse, making it highly competitive against existing methods.
The irregular impulses that occur before 2~ms cannot be fully reconstructed.
This may be because we use the direct-path speech in Eq.~(\ref{eq:lctf}) to get an aligned solution.
Since direct-path RIR deviates from an ideal Dirac delta and inherently contains part of the frequency response, the estimated RIR may lack the corresponding components. 
However, it is not a major limitation because practical applications typically do not rely on the very early response near the direct-path impulse \cite{wen2008blind,8937087,10096164}.

\begin{figure}[ht!]
    \centering
    \includegraphics[width=0.55\linewidth]{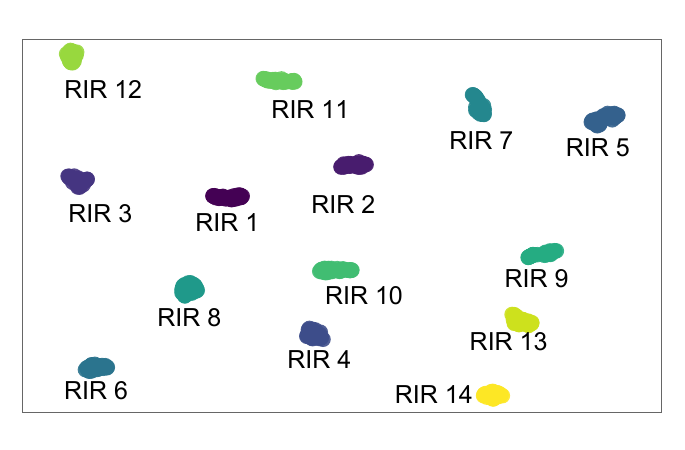}
    \vspace{-0.5em}
    \caption{2-D projections of CTF embeddings on SimACE.}
    \vspace{-1em}
    \label{fig:embeddings}
\end{figure}

Fig.~\ref{fig:embeddings} presents the 2-D projections of the CTF embeddings $\mathbf{e}_\mathrm{h}$ on the test set using UMAP~\cite{mcinnes2018umap} toolbox.
As observed, the embeddings exhibit a clear clustering corresponding to the RIR IDs, indicating that the proposed network effectively extracts spatial information from the observations.

\subsection{Ablation study}
\label{sec:ablation_study}

Table~\ref{tab:ablation} presents the metrics under different loss settings. 
The adopted setting offers significant advantages in DRR estimation.
Although the accuracy of T60 estimation is not optimal under the adopted setting, the statistical difference between it and the best result is not statistically significant (p-value $\approx 0.306 > 0.05$). 
In summary, the ablation study demonstrates that the proposed loss function is appropriate, and all auxiliary losses exert positive effects on the final performance.


\begin{table}[h!]
    \centering
    \vspace{-0.5em}
    \caption{Ablation study for loss settings}
    \vspace{-1em}
    \resizebox{0.99\linewidth}{!}{
    \begin{tabular}{c|c|c|c|c}
    \Xhline{1pt}
        Loss type&\multicolumn{2}{c|}{Loss weights}&RT60 (s)& DRR (dB)\\
        \Xcline{2-5}{0.4pt}
        in Eq. (\ref{eq:loss_x},\ref{eq:loss_s},\ref{eq:lctf})&$\lambda_\mathrm{denoi}$ &$\lambda_\mathrm{dereverb}$&MAE$\downarrow$&MAE$\downarrow$\\
        \Xhline{0.4pt}
        
        Mean square error&1.0&1.0&0.104$\pm$0.123&1.050$\pm$0.523\\
        Mag+RI (prop.)&0.0&0.0&0.077$\pm$0.080&1.056$\pm$0.505\\
        Mag+RI (prop.)&1.0&0.0&0.069$\pm$0.074&0.899$\pm$0.582\\
        Mag+RI (prop.)&0.0&1.0&\textbf{0.065}$\pm$\textbf{0.062}&0.960$\pm$0.394 \\
        \Xhline{0.4pt}
        \textbf{Mag+RI (prop.)}&\textbf{1.0}&\textbf{1.0} & 0.069$\pm$0.078&\textbf{0.684}$\pm$\textbf{0.402}\\
    \Xhline{1pt}
    \end{tabular}
    }
    \vspace{-1em}
    \label{tab:ablation}
\end{table}

\section{Conclusion}
\label{sec:conclusion}
We propose a blind RIR identification method via reverberant spectrum reconstruction, named Rec-RIR.
Rec-RIR proposes a multi-task DNN that sequentially removes noise and reverberation from observation and fuses reverberant speech embeddings with clean speech embeddings to realize end-to-end estimation of the CTF filter.
Experimental results demonstrate that Rec-RIR achieves SOTA performance in blind RIR identification.
\section{Generative AI Use Disclosure}
During the preparation of this work, the authors used ChatGPT to assist with language polishing.
The authors take full responsibility for the final manuscript.

\bibliographystyle{IEEEtran}
\bibliography{refs}

\end{document}